\documentclass[twocolumn]{svjour3}         % twocolumn
\smartqed  % flush right qed marks, e.g. at end of proof
\usepackage{graphicx}
\usepackage{subfig}
\newcommand{\pb}{$\bar{p}$}
\newcommand{\hmol}{H$_2$}

 \journalname{Hyperfine Interactions}
\begin{document}

\title{Collisions of low-energy antiprotons with molecular
  hydrogen: ionization, excitation and stopping power.%
%\thanks{Grants or other 
%notes 
%about the article that should go on the front page should be
%placed here. General acknowledgments should be placed at the end of the
%article.} 
}
%\subtitle{Do you have a subtitle?\\ If so, write it here}

\titlerunning{Collisions of low-energy antiprotons with molecular hydrogen}
% if too long for running head 

\author{Armin L\"uhr   \and
        Alejandro Saenz %etc.
}
%
%\authorrunning{Short form of author list} % if too long for running head
%
\institute{%Armin L\"uhr \at
              Institut f\"ur Physik, AG Moderne Optik, Humboldt-Universit\"at
              zu Berlin, Hausvogteiplatz 5-7, D-10117 Berlin, Germany. \\
              Tel.: +49-(0)30-2093-4814\\
%              Fax:  +49-(0)30-2093-4718\\
              \email{Armin.Luehr@physik.hu-berlin.de}            % \\
%             \emph{Present address:} of F. Author  %  if needed
           %\and
            %Alejandro Saenz \at
             % Institut f\"ur Physik, AG Moderne Optik, Humboldt-Universit\"at
              %zu Berlin, Hausvogteiplatz 5-7, D-10117 Berlin, Germany. \\
              %Tel.: +49-(0)30-2093-4902\\
%              Fax:  +49-(0)30-2093-4718\\
              %\email{Alejandro.Saenz@physik.hu-berlin.de}           % 
}

\date{Received: date / Accepted: date}
% The correct dates will be entered by the editor

\maketitle

\begin{abstract}
A time-dependent coupled-channel approach was used to calculate ionization,
excitation, and energy-loss cross sections as well as energy spectra for
antiproton and proton collisions with molecular hydrogen for impact energies
8\,$<$\,$E$\,$<$\,4000 keV.% using a model potential.
% \PACS{PACS code1 \and PACS code2 \and more}
% \subclass{MSC code1 \and MSC code2 \and more}
\end{abstract}
%
%
%%%%%%%%%%%%%%%%%%%%%%%%%%%%%%%%%%%%%%%%%%%%%%%%%%%%%%%%%
%
\section{Introduction}
\label{sec:intro}
A large amount of theoretical work has been done for antiproton (\pb )
collisions with He atoms stimulated by discrepancies between experiment and
theory lasting for more than a decade~\cite{anti:knud08}. In the case of
hydrogen also a number of contributions exist for atomic targets whereas the
knowledge on low-energy \pb\ collisions with the simplest two-electron
molecule \hmol\ is rather poor \cite{anti:ande90a,anti:hvel94}, especially,
concerning theory \cite{anti:ermo93,anti:luhr08a}. Precise data on
\pb\ +\,\hmol\ are, however, of great interest in many fields. They can, e.g.,
be used to determine the stopping power which  is a prerequisite for the
design of low-energy \pb\ storage rings taking the interactions with
residual-gas atoms and molecules into account. But also for the preparation of
accurate future experiments with low-energy \pb , which in turn can be used
as a test of competing theoretical approaches, the maximum of the
stopping power is of importance. 
A short overview on the present status of  \pb\,+\,\hmol\ shall be given.

The used method has already been discussed in some detail elsewhere
\cite{anti:luhr08,anti:luhr08a,dia:luhr08} as well as partly in the article on
interactions of antiprotons with alkali-metal atoms \cite{anti:luhr09a}.
A time-dependent close-coupling approach is employed which
uses the 
impact-parameter method. The target molecule is treated as an effective
one-electron system by applying a model potential
\cite{dia:luhr08,sfm:vann08}. The wave function of the collision  
process is expanded in a one-center approach in eigenfunctions of the
one-electron model Hamiltonian of the target. Thereby, the radial part is
expanded in B-spline functions and the angular part in a 
symmetry-adapted sum of spherical harmonics.

%
%
%%%%%%%%%%%%%%%%%%%%%%%%%%%%%%%%%%%%%%%%%%%%%%%%%%%%%%%%%%
%
\section{Results}
\label{sec:results}
%
%
%
%
%%%%%%%%%%%%%%%%%%%%%%%%%%
%
%\subsection{Ionization}
\paragraph{Ionization}
\captionsetup[subfloat]{labelfont={bf},labelformat=simple}
\begin{figure}
\centering
  \subfloat[%
  Ionization cross section.
  Theory.  
  \pb : solid curve, present; dashed curve, Ermolaev; dotted curve,
  two  times atomic H. 
  $p$: dash--dotted curve, present.
  Experiment. \pb \,:
  squares, Hvelplund {\it et al.}~\cite{anti:hvel94};
  circles, Anderson {\it et al.}~\cite{anti:ande90a}.
  $p$\,:    
  triangles, Rudd {\it et al.}~\cite{sct:rudd83};%,sct:rudd85};
  diamonds, Shah {\it et al.}~\cite{sct:shah82}.%,sct:shah89}.
  \label{fig:ionization}] 
  {\includegraphics[width=0.45\textwidth]{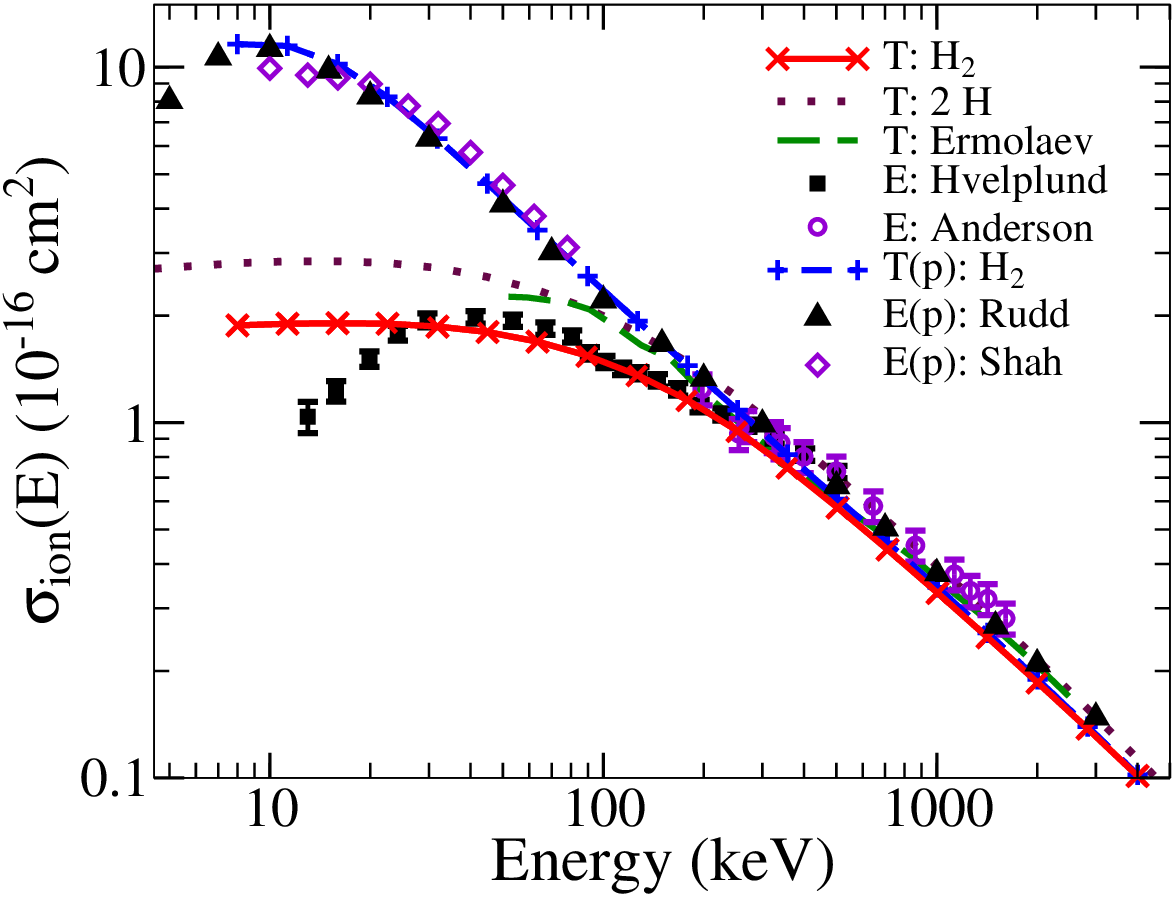}} 
  \hspace{0.03\textwidth}
  \subfloat[%
  Excitation cross section.
  Theory. \hmol : solid curve, total; dashed curve, p states; dash--dotted
  curve, 2p.  
  H: dash--doubly-dotted curve, total; dotted curve, two  times total. 
  Experiment. \hmol : 
  squares, $e^-$: X $\rightarrow$ B\,+\,C, Liu {\it et al.}~\cite{sct:liu98}. 
  (See text)
  \label{fig:excitation}]{
    \includegraphics[width=0.45\textwidth]{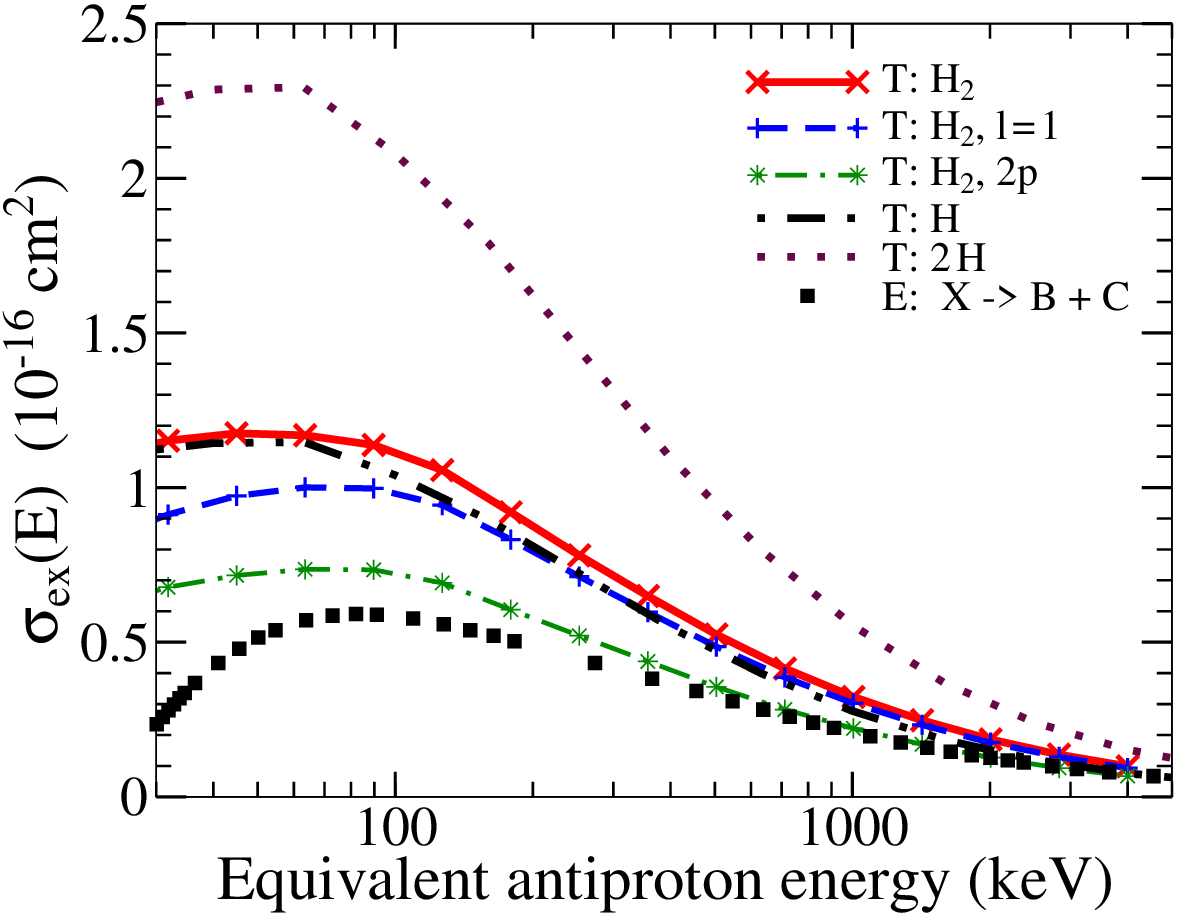}
  }  
\end{figure}
\addtocounter{subfigure}{2}
In Fig.~1 the present \hmol\ results for ionization and electron loss by
\pb\ and $p$ impact, respectively, are compared with cross sections given in
the literature and results for {\it atomic} hydrogen multiplied by two. It can
be seen that the naive picture of 
the \hmol\ molecule as being practically the same as two independent H atoms
only holds at high impact energies $E$. For high and intermediate impact
energies the present \pb\ results agree with the experiments by Anderson {\it et
  al.}~\cite{anti:ande90a} and Hvelplund {\it et al.}~\cite{anti:hvel94}
although they are slightly lower than the measured data. At low energies
$E<20$\,keV, however, 
the experimental results are significantly smaller than the present
calculations which partly can be explained with experimental problems not
accounted for  at that time \cite{anti:knud08a}.
These problems also occurred for \pb\ + He cross
sections measured with the same apparatus \cite{anti:hvel94} and were
identified by very recent He measurement~\cite{anti:knud08}. The
other theoretical curve~\cite{anti:ermo93} describes the 
\pb\ + \hmol\ cross section only for $E>200$\,keV satisfyingly. For smaller $E$,
however, it follows rather the curve for twice the hydrogen atom.  This
behavior may originate from an insufficient description of the continuum
(cf.\ \cite{anti:luhr08a}). 

For high energies the present cross sections for $p$ and \pb\ are practically
the same as can be expected in the validity regime of the first Born
approximation which is independent of the sign of the projectile charge. 
For these energies the experimental results for \pb\ are, however,
larger than those for $p$ by Rudd {\it et al}. This may originate from
different normalization procedures. On the other hand, the present $p$ results
agree with the experimental data in the whole energy range considered
here; confirming the applicability of the used method even for $p$ collisions.

\paragraph{Excitation}
The results for excitation are presented in Fig.~2. A
comparison of \emph{molecular} with \emph{atomic} hydrogen shows clearly that
the simplified picture of \hmol\ behaving basically like 
two H atoms is not senseful in the present context. Remarkably, the excitation
cross section for \emph{one} H atom is practically the same as for \hmol . It
can also be seen that transitions into p ($l=1$) states (of the model
potential) are by far the most prominent excitations. In particular, the
differential cross section for the first excited 2p state is of the order of
2/3 of the total excitation cross section. In the case of \hmol\ excitation by
\pb\ neither experimental nor other theoretical results exist in the
literature to the authors' knowledge. Therefore, the present results for the
excitation into the 2p state are compared in Fig.~2 with experimental
electron-collision results by Liu {\it et al.}~\cite{sct:liu98} for
excitations from the \hmol\ ground state X into the energetically-lowest
dipole-allowed states labeled with B and C  
which are orientationally averaged \cite{dia:luhr08}. The present results
for \pb\ impact agree nicely with the experiment as expected at high
energies. However, experimental excitation data for \pb\ impact would be
valuable in order to judge the quality of the results at energies below the
maximum at $E\approx50$ keV. 
%
%%%%%%%%%%%%%%%%%%%%%%%%%%
%
\paragraph{Electron-energy spectra}
\begin{figure}[t]
\centering
  \subfloat[Electron-energy spectra  for $E$\,=\,48\,\,keV. 
  Theory: solid  curve, \pb ; dashed curve, $p$. 
  Experiment: dots, Gealy {\it et\,\,al.} \cite{sct:geal95a}.%
  \label{fig:spectrum_a}]{%     
    \includegraphics[width=0.45\textwidth]{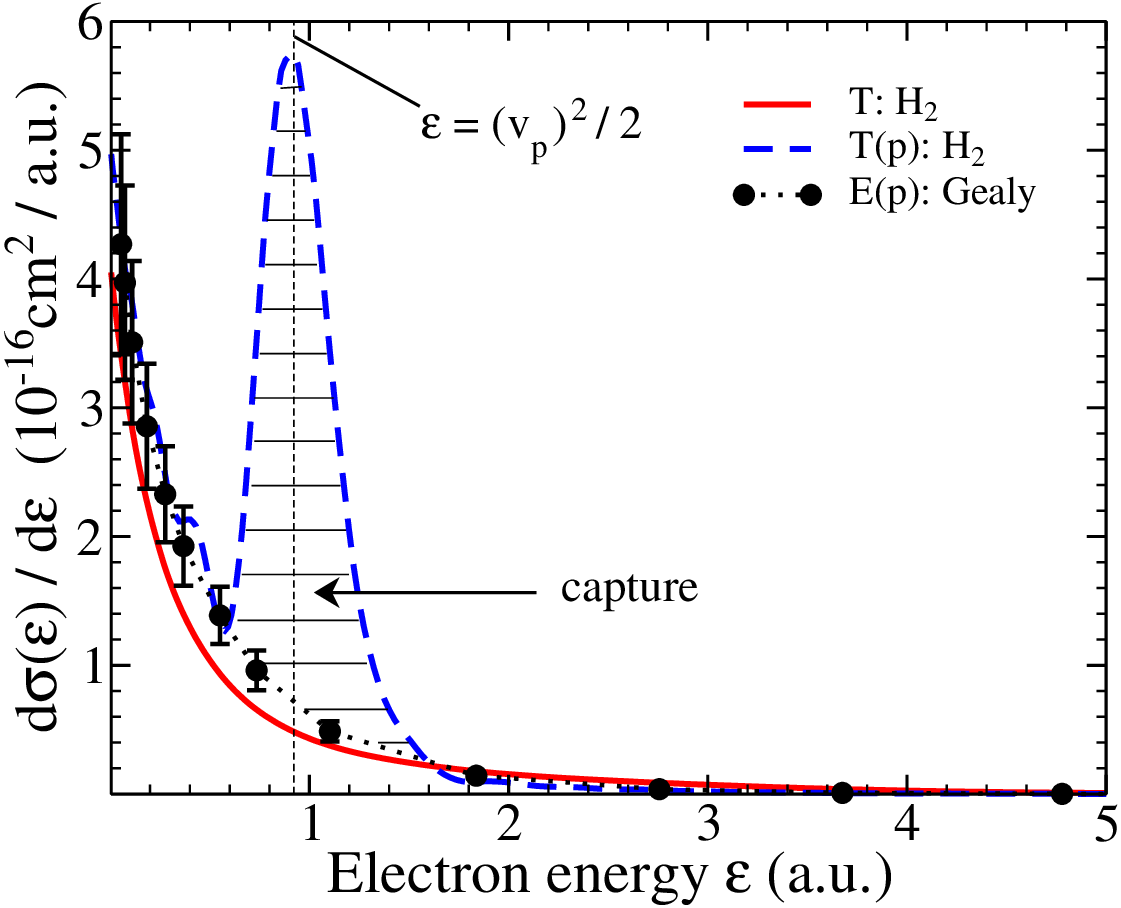}
  }  
   \hspace{0.03\textwidth}
   \subfloat[Electronic stopping power for \hmol . 
   Theory: solid curve, \pb . 
   Experiment: dashed curve, \pb , Lodi Rizzini {\it et al.}~\cite{anti:lodi02};
   dash--dotted curve, $\mu^-$, Hauser {\it et al.}~\cite{anti:haus93}. % 
   \label{fig:spectrum_c}]{%   
     \includegraphics[width=0.45\textwidth]{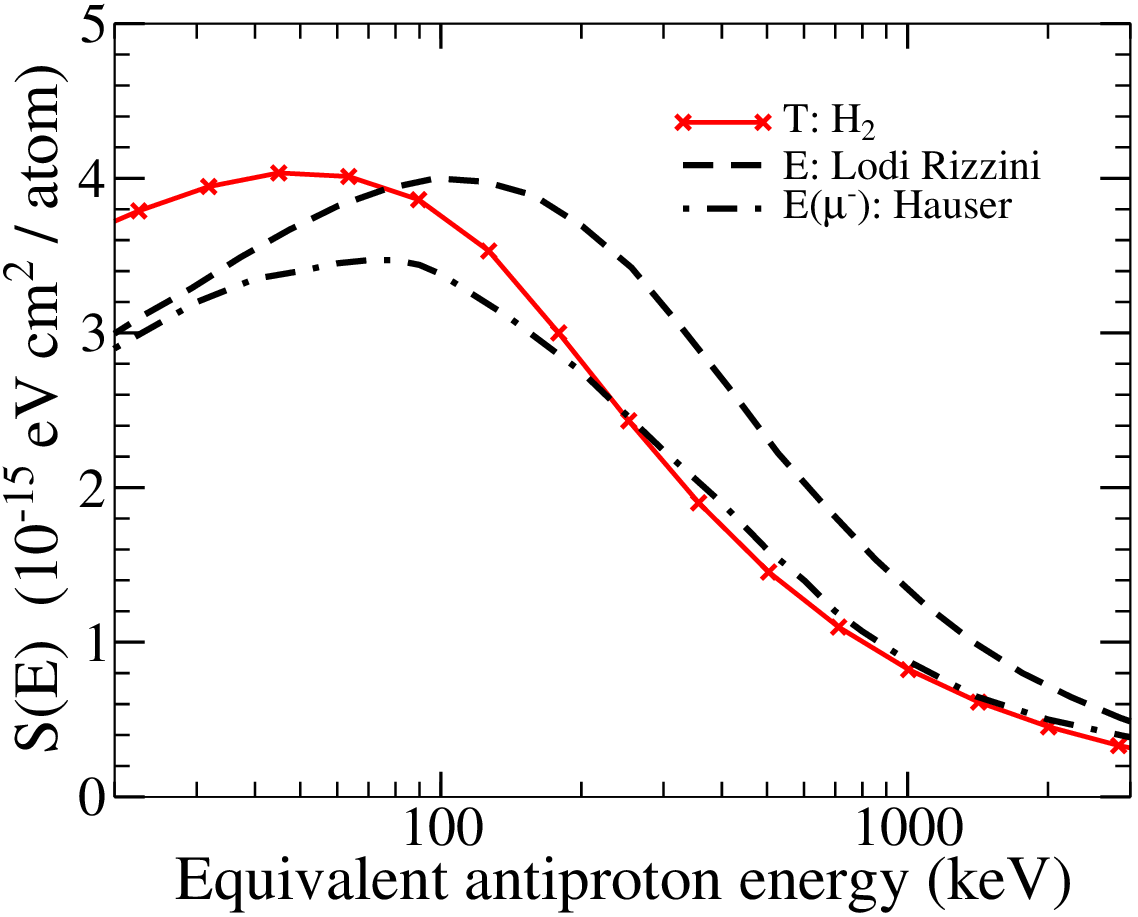}
   } 
\end{figure}
In Fig.~3 electron-energy spectra for \pb\ and $p$ collisions
with \hmol\ are presented for $E=48$ keV. The comparison of the \pb\ and $p$
calculations reveals different slopes for both curves.
The most striking
difference, however, is the peak in the case of $p$ impact at an electron
energy $\epsilon$ which corresponds to the velocity of the proton whereas the
\pb\ curve is lowered around this $\epsilon$. A comparison of the $p$ curve
with measured data by Gealy {\it et al.}~\cite{sct:geal95a} yields good
agreement except for the mentioned peak. The area under the maximum (hatched)
and therefore the difference between the present and the experimental %$p$
curve corresponds to the cross section for electron capture which is excluded
in the experimental data. 
%
%%%%%%%%%%%%%%%%%%%%%%%%%%
%
%
%
%
\paragraph{Stopping power}
Using the differential information on ionization and excitation discussed so
far it is possible to determine the electronic energy-loss cross section also
referred to as stopping power $S$. In Fig.~4 the present results for \pb\ +
\hmol\  are compared with experimental data by Lodi Rizzini {\it et
  al.}~\cite{anti:lodi02}. Also, data for negative muon ($\mu^-$) impact is
given measured by Hauser {\it et al.}~\cite{anti:haus93}. Except for
deviations at small velocities due to the mass dependence of the nuclear
stopping power the results for \pb\ and $\mu^-$ should be the same. The
obvious discrepancies ($\approx\,20\,\%$) between \pb\ and $\mu^-$ indicate
the experimental difficulties. The present results are for $E>200$ keV in good
agreement with the  $\mu^-$ data. The ionization cross sections
\cite{anti:ande90a} used as input for the analysis in \cite{anti:haus93} were,
however, found to be erroneous for $E<200$ keV
\cite{anti:hvel94,anti:knud08a}. On the other hand, the slope of the present
curve for $E>100$ keV is remarkably similar to that of the experimental \pb\
curve but for $E>160$ keV. This means that there seems to be a scaling of $E$
by a factor of $\approx$1.6 between the present and the experimental \pb\ data.
The stopping power for \pb\ + \hmol\ is currently investigated in detail in
order to clarify this issue.

\section{Conclusion}
The present results obtained with a model 
potential are meaningful for ionization, excitation, and even for differential
excitation and electron spectra for energies above the maximum around 60
keV. This opens up the possibility to further analyze the data in order to
obtain doubly-differential cross sections. But also ---as has been done
here--- to extract quantities like the stopping power for  \pb\ + \hmol\
collisions.  

The differences for ionization between experiment and theory at low \pb\
energies as well as between experiments for \pb\ and $p$ impact at
high energies and the lack of measurements of excitation 
obviously show the need for new experimental \pb\,+\,\hmol\ data which may
be available in the near future \cite{anti:knud08a}. 
However, also an improvement of the theoretical description is necessary to
account for the expected increase of importance of molecular (e.g., charge
asymmetry, dissociation) and two-electron effects for low-energy
impacts $E<100$ keV which are already indicated by the present
findings. 
Additionally, the present stopping power results for $E>100$ keV agree with the
experimental data for $\mu^-$ but disagree for \pb\ making further
investigation of $S(E)$ necessary.  

\begin{acknowledgements}
The authors are pleased to acknowledge discussions of the experimental results
with Professors H.~Knudsen and F.~Kottmann.
This work was supported by BMBF (FLAIR Horizon) and {\it Stifterverband f\"ur
  die deutsche Wissenschaft}.
\end{acknowledgements}

% BibTeX users please use one of
%\bibliographystyle{spbasic}      % basic style, author-year citations
\bibliographystyle{spmpsci}      % mathematics and physical sciences
%\bibliographystyle{spphys}       % APS-like style for physics
%\bibliography{anti,sct,sfm,dia}   % name your BibTeX data base

% Non-BibTeX users please use

\end{document}